\documentclass{emulateapj}
\usepackage[utf8]{inputenc}
\usepackage{apjfonts}
\usepackage{graphicx}
\usepackage{color}

\usepackage{epstopdf}
\usepackage{amsmath}
\newcommand\lsim{\mathrel{\rlap{\lower4pt\hbox{\hskip1pt$\sim$}}
\raise1pt\hbox{$<$}}}
\newcommand\gsim{\mathrel{\rlap{\lower4pt\hbox{\hskip1pt$\sim$}}
\raise1pt\hbox{$>$}}}

\begin{document}
\title{Hidden Planetary Friends: On the stability of 2-planet systems in the presence of a distant, inclined companion }
\author{Paul Denham$^{1}$,  Smadar Naoz$^{1,2}$, Bao-Minh Hoang$^{1}$, Alexander P.~Stephan$^{1}$, Will M.~Farr$^{3}$ }
\affil{ $^{1}$Physics and Astronomy Department, University of California, Los Angeles, CA 90024} 
\affil{$^{2}$Mani L. Bhaumik Institute for Theoretical Physics, Department of Physics and Astronomy, UCLA, Los Angeles, CA 90095, USA}
\affil{$^{3}$School of Physics and Astronomy, University of Birmingham, Birmingham, B15 2TT, UK}

\email{pdenham629@gmail.com \\ snaoz@astro.ucla.edu}
\begin{abstract}
Recent observational campaigns have shown that multi-planet systems seem to be abundant in our Galaxy. Moreover, it seems that these systems might have distant companions, either planets, brown-dwarfs or other stellar objects. These companions might be inclined with respect to the inner planets, and could potentially excite the eccentricities of the inner planets through the Eccentric Kozai-Lidov mechanism. These eccentricity excitations could perhaps cause the inner orbits to cross, disrupting the inner system. We study the stability of two-planet systems in the presence of a distant, inclined, giant planet. Specifically, we derive a stability criterion, which depends on the companion's separation and eccentricity. We show that our analytic criterion agrees with the results obtained from numerically integrating an ensemble of systems. Finally, as a potential proof-of-concept, we provide a set of predictions for the parameter space that allows the existence of planetary companions for the Kepler-56, Kepler-448, Kepler-88, Kepler-109, and Kepler-36 systems.
\end{abstract}

\section{Introduction}

Recent ground and space-based observations have shown that multi-planet systems are abundant around main-sequence stars \citep[e.g.,][]{Howard+10,Howard+12,Borucki+11,Lissauer+11,Mayor+11,Youdin11,Batalha+13,Dressing+13,Petigura+13,Christiansen+15}. These studies reveal that the architecture of planetary systems can drastically vary from our solar system. For example, systems consisting of multiple low mass (sub-Jovian) planets with relatively compact orbits usually have periods that are shorter than Mercury's.

NASA's Kepler mission found an abundance of compact multi-planet super-Earths or sub-Neptune systems \citep[e.g.,][]{Mullally+15,Burke+15,Morton+16,Hansen17}.  These systems seemed to have low eccentricities \citep[e.g.,][]{Lithwick+12,Van+15}. In addition, it was suggested that these many body systems are close to the dynamically stable limit \citep[e.g.,][]{Fang+13,Pu+15,Volk+15}. It was also proposed that single planet systems might be the product of systems initially consisting of multiple planets that underwent a period of disruption, i.e, collisions, resulting in reducing the number of planets (see \citet{Johansen+12} \citet{Becker+16}).


Giant planets may play a key role in forming inner planetary systems. Radial velocity surveys, along with the existence of Jupiter, have shown giants usually reside at larger distances from their host star than other planets in the system ($\geq 1$~AU)\citep[e.g.,][]{Knutson+14K,Bryan+16}. For example, there were some suggestions that the near-resonant gravitational interactions between giant outer planets and smaller inner planets can shape the configuration of an inner system's asteroid belt \citep[e.g.,][]{Williams+81,Minton+11}. Hence, it was suggested that our solar system's inner planets are of a second generation, which came after Jupiter migrated inward to its current orbit \citep[e.g.,][]{Batygin+15}.  Furthermore, secular resonance may be the cause of water being delivered to earth \citep{1986A&A...170..138S,1994P&SS...42..301M}, the potential instability of Mercury's orbit \citep{1997A&A...318..975N,2009sf2a.conf..105F,2008ApJ...683.1207B,2011ApJ...739...31L}, and the obliquity of the exoplanets' in general \citep[e.g.,][]{Naoz11,Li+14Kepler}.

Recently, several studies showed that the presence of a giant planet can affect the ability to detect an inner system \citep[e.g.,][]{Hansen17,Becker+17,Mustill+17,Jontof+17,Huang+17}. Specifically, dynamical interactions from a giant planet, having a semi-major axis much greater than the planets in the inner system, can excite eccentricities and inclinations of the inner planets. A possible effect is that the inner system becomes incapable of having multiple transits or completely unstable. \citet{Volk+15} showed that observed multi-planet systems may actually be the remnants of a compact system that was tighter in the past but lost planets through dynamical instabilities and collisions \citep[see also][]{Petrovich+14}. Interestingly, verifying these problems could reconcile the Kepler dichotomy \citep[e.g.,][]{Johansen+12,Ballard+16,Hansen17}. 


In this work, we investigate the stability of compact sub-Jovian inner planetary systems in the presence of a distant giant planet (see Figure \ref{fig:sailing} for an illustration of the system). Over long time scales, a distant giant planets' gravitational perturbations can excite the eccentricities of the inner planets' to high values, destabilizing the inner system. \citep[e.g.,][]{Naoz11}. However, if the frequency of angular momentum exchange between the inner planets is sufficiently high, then the inner system can stabilize. Below we derive an analytic stability criterion (Section \ref{sec:stab}). Then we analyze our nominal system in Section \ref{sec:Long} and provide specific predictions for Kepler systems in Section \ref{sec:Kepler}. Finally, we offer our conclusion in Section \ref{sec:dis}. 

Near the completion of this work, we became aware of \citep{Pu+18} a complementary study of stability in similar systems. Here we provide a comprehensive stability criterion as a function of the companion's parameters. Furthermore, we provide a set of predictions for possible hidden companion orbits in several observed systems.

\begin{figure}
	\centering
		\includegraphics[width=\linewidth]{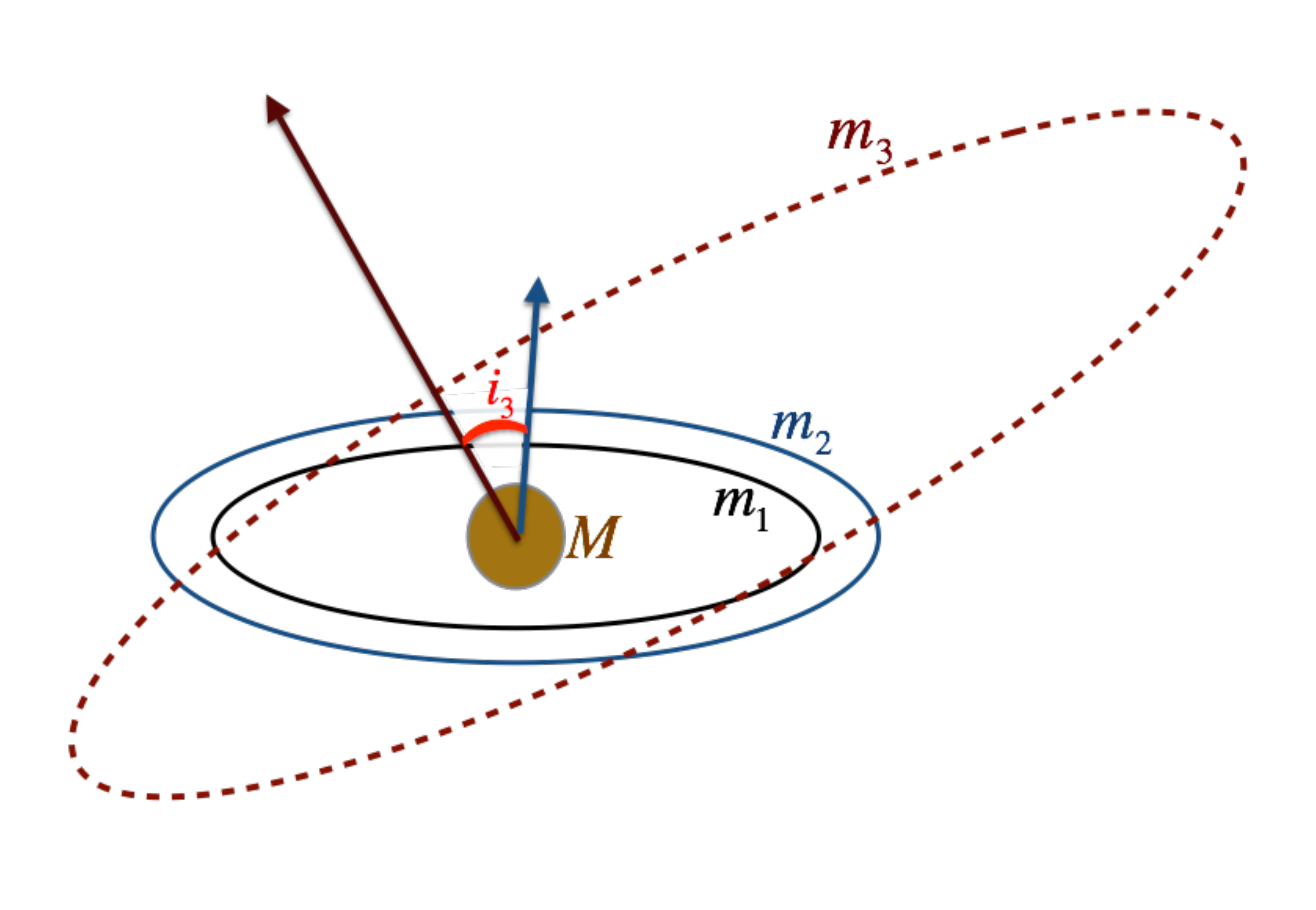}
		
	\caption{Here is a schematic of the systems being analyzed. The inclination of the third planet, $i_3$, is shown and measured relative to the z-axis, along the spin axis of the host star. $M, m_1, m_2$, and $m_3$ are the masses of the host star, and the first, second and third planets respectively. }
	\label{fig:Catroon}
\end{figure}

\section{Analytical Stability criterion  }\label{sec:stab}

Here we develop a generic treatment of the stability of two-body systems in the presence of an inclined outer planet. 
Consider an inner system consisting of two planets ($m_1$ and $m_2$) orbiting around a host star $M$ with a relatively tight configuration (with semi-major axis $a_1$ and $a_2$, respectively). We introduce an inclined and eccentric companion that is much farther from the host star than the planets in the inner system ($m_3$ and $a_3$, see Figure \ref{fig:Catroon} for an illustration  of the system). We initialize this system to have orbits far from mean-motion resonance, and have $a_1,a_2<<a_3$. The three planets' orbits have corresponding eccentricities  $e_1$, $e_2$ and $e_3$.  We set an arbitrary $z$-axis to be perpendicular to the initial orbital plane of the inner two planets, thus, the inclinations of $m_1$, $m_2$ and $m_3$ are defined as $i_1$, $i_2$ and $i_3$. Accordingly, for planets one, two, and three we denote the longitude of ascending nodes and the argument of periapse $\Omega_1$, $\Omega_2$, $\Omega_3$, $\omega_1, \omega_2$, and $\omega_3$.

The outer orbit can excite the eccentricities, via the Eccentric Kozai-Lidov mechanism (EKL) (\citet{Kozai}, \citet{Lidov}, see \citet{Naoz16} for review) on each planet in the inner system. The EKL resonance causes angular momentum exchange between the outer and inner planets, which in turn causes precession of the periapse of each of the inner orbits. However, angular momentum exchange between the inner two planets also induces precession of the periapse \citep["the so-called Laplace-Lagrange interactions, e.g.,][]{MD00}. If the inner orbits' angular momentum exchange takes place at a faster rate than that induced by the outer companion, then the system will not be disrupted by perturbations from the tertiary planet. The quadrupole approximation to the timescale of the EKL between the third planet $m_3$ and the second one, $m_2$ is given by: \begin{equation}\label{eq:Kozai}
\tau_{k(2,3)} \sim \frac{16}{15}\frac{a_3^3}{a_2^{3/2}} \frac{\sqrt{M+m_2+m_1}}{m_3\sqrt{G}}(1-e_3^2)^{3/2} \ ,
\end{equation}
\citep[e.g.,][]{Antognini15}
where $G$ is the gravitational constant. This is roughly the timescale at which the second planet's argument of periapse precesses.  Note that here we considered the timescale of EKL excitations from the tertiary on $m_2$ since this timescale is shorter than the timescale of EKL excitations between $m_3$ and $m_1$. Planet $m_2$'s argument of periapse also precesses due to gravitational perturbations from the inner-most planet, $m_1$. The associated timescale is \citep[e.g.,][]{MD00}
\begin{eqnarray}
\tau_{LL}&\sim &\bigg[A_{22}+A_{21}\left(\frac{e_1}{e_2}\right)\cos(\varpi_2-\varpi_1)\nonumber \\ &-& B_{22}-B_{21}\left(\frac{i_1}{i_2}\right)\cos(\Omega_2-\Omega_1)\bigg]^{-1} \ ,
\end{eqnarray}
where $\varpi_j=\omega_j+\Omega_j$ for $j=1,2$. The $A_{i,j}$, and $B_{i,j}$ are laplace coefficients, which are determined by:
\begin{eqnarray}
A_{22} &=& n_2\frac{1}{4\pi}\frac{m_1}{M+m_2}\left(\frac{a_1}{a_2}\right)^2 f_{\psi} \ , \\
A_{21}&=&-n_2\frac{1}{4\pi}\frac{m_1}{M+m_2}\left(\frac{a_1}{a_2}\right) f_{2\psi} \ , \\
B_{22}&=&-n_2\frac{1}{4\pi}\frac{m_1}{M+m_2}\left(\frac{a_1}{a_2}\right)^2  f_{\psi} \ , \\
B_{21} &=& n_2\frac{1}{4\pi}\frac{m_1}{M+m_2}\left(\frac{a_1}{a_2}\right)f_{\psi}\ . \end{eqnarray}
and
\begin{eqnarray}
f_{\psi} &=& \int_{0}^{2\pi}\frac{\cos\psi}{ \left(1-2\left(\frac{a_1}{a_2}\right)\cos\psi+\left(\frac{a_1}{a_2}\right)^2\right)^\frac{3}{2} }d\psi  \ , \\
f_{2\psi} &=&\int_{0}^{2\pi}\frac{\cos2\psi}{\left(1-2\left(\frac{a_1}{a_2}\right)\cos\psi+\left(\frac{a_1}{a_2}\right)^2\right)^\frac{3}{2}} d\psi \ .
\end{eqnarray}

The system will remain stable if angular momentum exchange between the two inner planets takes place faster than the precession induced by $m_3$. Accordingly, there exists two regions of parameter space: one contains systems where perturbations on $m_2$ are dominantly from $m_1$, and the other contains systems where $m_2$ is dominated by $m_3$. The former is called the Laplace-Lagrange region, and the ladder is called EKL region. The transition between the Laplace-Lagrange region and EKL region is determined by comparing the two timescales relating the frequencies of precession from each mechanism, i.e.,
\begin{equation}\label{eq:timescales}
    \tau_k \sim \tau_{LL} \ .
\end{equation}
 We have related the timescale for Kozai oscillations induced by $m_3$ on $m_2$ to the Laplace-Lagrange timescale between $m_1$ and $m_2$. Equating these timescales yields a simple expression for the critical eccentricity of the third planet, $e_{3,c}$, as a function of $a_2$, i.e.,
\begin{equation}\label{eq:e3c}
   e_{3,c} \sim\left(1-\left[\frac{15}{16}\frac{m_3\sqrt{G}}{\sqrt{M_\odot+m_1+m_2}}\frac{a_2^{3/2}}{a_3^3}\frac{1}{f_{LL} }\right]^{\frac{2}{3}}\right)^{\frac{1}{2}} \ ,
\end{equation}
where 
\begin{equation}\label{eq:fLL}
f_{LL}=A_{22}+A_{21} \frac{e_{1,ic}}{ e_{2,ic}}\cos(\varpi_1-\varpi_2)-B_{21}\frac{i_{1,ic}}{i_{2,ic}}\cos(\Omega_1-\Omega_2) -B_{22} \ .
\end{equation}
$e_{j,ic}$ and $i_{j,ic}$, are the initial eccentricity and inclination for the two inner planets, i.e., $j=1,2$. 
 As we show below, during the evolution of a stable system $\Omega_1\sim\Omega_2$, $\varpi_2\sim \varpi_1$, $e_1\sim e_2$, and $i_1\sim i_2$. Thus, we define a minimum stable configuration of $f_{LL,{\rm min}}$
 \begin{equation}\label{eq:fLLmin}
f_{LL,{\rm min}}=A_{22}+\frac{e_{1,ic}}{ e_{2,ic}}A_{21} - \frac{i_{1,ic}}{i_{2,ic}}B_{21} -B_{22} \ .
\end{equation}
 Numerically we find that during the evolution of an unstable system, $e_1/e2$, $i_1/i_2$, $\Omega_1/\Omega_2$, and $\varpi_2/ \varpi_1$ largely deviate from unity, where $\cos(\varpi_1-\varpi_2)$ can be negative. Thus, we also define the maximum stability as:
  \begin{equation}\label{eq:fLLmax}
f_{LL,{\rm max}}=A_{22}-\frac{e_{1,ic}}{ e_{2,ic}}A_{21} -\frac{i_{1,ic}}{i_{2,ic}} B_{21} -B_{22} \ ,
\end{equation}
where the difference between Eq.~(\ref{eq:fLLmin}) and (\ref{eq:fLLmax}) is the sign of $A_{21}$. 
The stability of systems transitions from the minimum to the maximum $f_{LL}$, as a function of $e_3$. In other words we find a band of parameter space, between $e_{3,c}(f_{LL,{\rm min}})$ and $e_{3,c}(f_{LL,{\rm max}})$, where systems are nearly unstable or completely unstable. 
 If the third planet's eccentricity is larger than the right hand side of Equation (\ref{eq:e3c}), then the inner system is more likely to become unstable.  In the next section we test this stability criterion and show that it agrees with secular numerical integration.

 We note that this stability criterion is based on the Laplace-Lagrange approximation, which assumes small eccentricities and inclinations for orbits in the inner system. Thus, it might break down for initial large eccentricities or mutual inclinations of the inner two planets. Furthermore, this stability criterion assumes that the inner system is compact and that the eccentricity excitations from the second planet on the first are suppressed. On the other hand, in the presence of these conditions the stability criterion depends only on the shortest EKL timescale induced by the far away companion and the corresponding Laplace-Lagrange timescale. Thus, it is straightforward to generalize it to more than two inner planets.

\section{Long term stability of multi-planet system}\label{sec:Long}

\subsection{The Gaussian averaging method }

To test our analytic stability criterion we utilize Gauss's Method. This prescription allows us to integrate the system over a long timescale (set to be $10$~Gyr, see below), in a time-efficient way.  In this mathematical framework, the non-resonant planets' orbits are phase-averaged and are treated as massive, pliable wires interacting with each other. The line-density of each wire is inversely proportional to the orbital velocity of each planet.  The secular, orbit-averaged method requires that the semi-major axes of the wires are constants of motion. We calculate the forces that different wires exert on each other over time. 

 This method was recently extended to include softened gravitational interactions by introducing a small parameter in the interaction potentials to avoid possible divergences while integrating \citep{Touma+09}. Furthermore, the method has been proven to be very powerful in addressing different problems in astrophysics, ranging from understanding the dynamics in galactic nuclei to describing how debris disks evolve. \citep[e.g.,][]{Touma+09,Sridhar+16,Nesvold+16,Saillenfest+17}.

\subsection{Stability test on a nominal system}

We tested our stability criterion by numerically integrating an ensemble of systems with identical initial conditions except for the semi-major axis and eccentricity of the second and third planets respectively. In this ensemble, we initialize all systems with a primary star $M=1$~M$_\odot$ orbited by three planets where the two inner planets, $m_1$, and $m_2$, each have masses of $1$~M$_\oplus$. The innermost planet, $m_1$, was placed in an orbit with a semi-major axis of $a_1=1$ au, and both inner planets were set initially on circular orbits in the same orbital plane. We set the third planet with a mass of $m_3=1$~M$_J$, at $a_3=100$ au. Throughout the ensemble, $a_2$ ranges from $1.4$ au to $5.9$ au, in steps of $0.5$ au, and $e_3$ ranges from zero to $0.95$ in steps of $0.05$. In all such systems, the third planet's inclination was set to be $45^\circ$ relative to the inner system. We integrated each system to $10$~Gyr, or until orbits crossed. From the results we made a grid labeling the stable/unstable systems of the ensemble. The grid revealed that stable orbits are bounded by the stability criterion. The same process was done on an ensemble which had the third planet's inclination set to $65^\circ$ instead; the result was the same. 

\begin{figure}[!h!t]
	\centering
		\includegraphics[width=\linewidth]{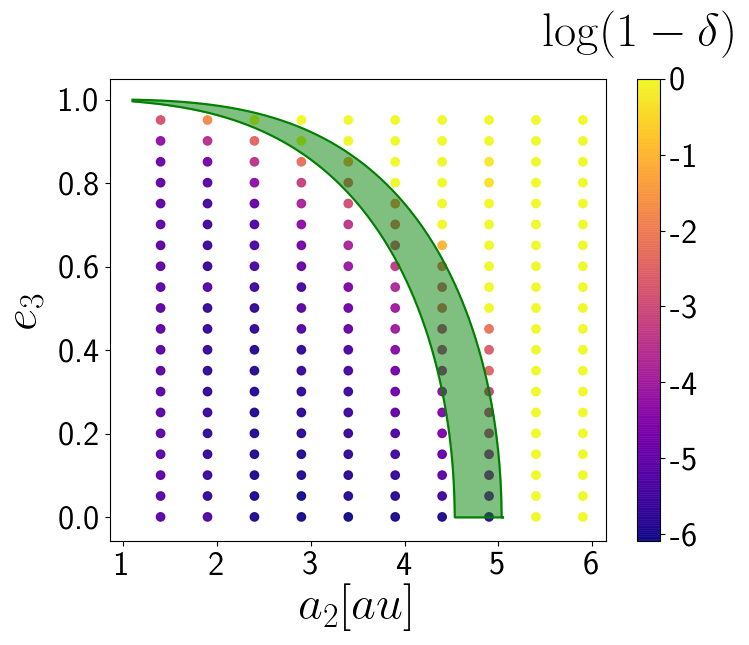}
			\caption{Here is the parameter space relating the third planets' eccentricity with the second planet' separation. The distant companion's inclination was set to $45^\circ$.
		The color code shows $\log(1-\delta)$, Eq.~(\ref{eq:delta}). This parameter estimates the closest the two inner orbits came to each other during the evolution. A darker color characterizes far away separation while a lighter color characterizes orbit crossing. 
		The stability criterion,  Eq.~(\ref{eq:e3c}), is plotted over the grid to show that stable orbits are bounded. The bottom green line represents $e_{3,c}(f_{LL,{\rm min}})$ and the top green line represents $e_{3,c}(f_{LL,{\rm max}})$, and thus the shaded band is the transition zone (see text).  Systems above the zone undergo an instability episode while systems below the zone are stable. }
	\label{fig:sailing}
\end{figure}

In Figure \ref{fig:sailing} we show the grid of systems we integrated in the parameter space relating $a_2$ and $e_3$. The color code is determined by a proxy that characterizes the stability of the system. The proxy is defined as $\log(1-\delta)$, where $\delta$ is a parameter which describes how close the two inner orbits came during the evolution: 
\begin{equation}\label{eq:delta}
    \delta={\rm min}\bigg [\frac{a_2(1+e_2) - a_1(1-e_1)}{a_2-a_1} \bigg ] \ .
\end{equation}
In Figure \ref{fig:sailing}, a lighter color means a smaller value of $\delta$, which when zero yields orbit crossing, while a darker color represents a stable configuration. 
The criterion for stability is plotted over the grid to show that it is in agreement with the probability for orbits to cross. Above the stability curve, systems are more likely to destabilize during evolution.

\begin{figure*}
\centering
\includegraphics[scale=.9]{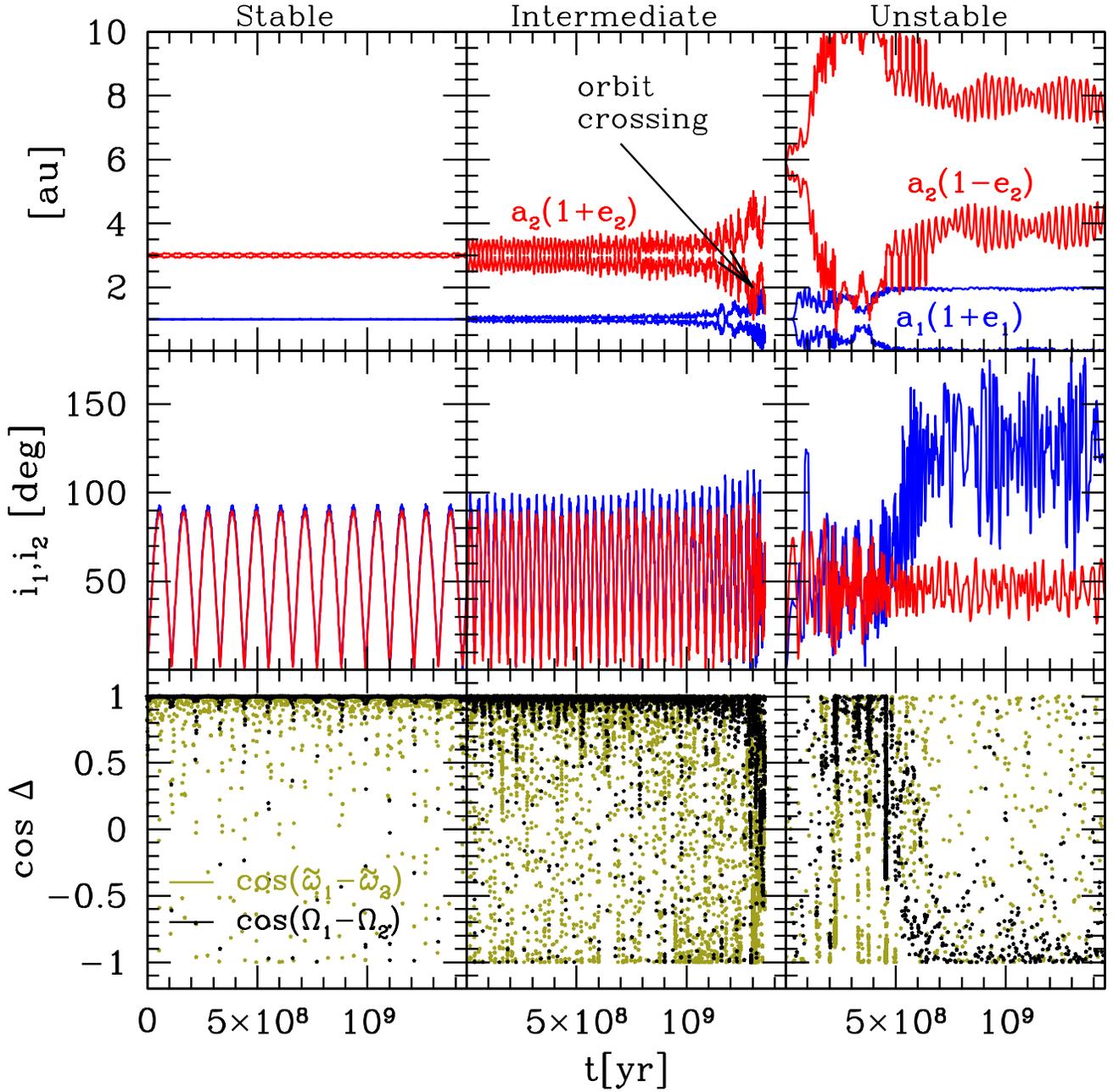}\caption{ \upshape Here are the three prominent types of dynamics from each region. From left to right, stable, intermediate, and unstable, for our nominal system. Shown in the top panels are the apocenters and the pericenters of the innermost planets ($m_1$ in blue and $m_2$) in red. The middle panels show the inclinations of the inner planets ($i_1$, in blue and $i_2$, in red). {\it Note that this inclination is not with respect to the total angular momentum, but rather with respect to the initial angular momentum of the two inner planets.} The bottom panels show, colored in black, the difference between the longitude of ascending nodes of the two inner planets as $\cos (\Omega_1-\Omega_2)$, and, in yellow, the difference between the longitude of the periapsis as $\cos (\varpi_1-\varpi_2)$. These two parameters are present in Equation (\ref{eq:fLL}). 
Recall that the nominal system has the following parameters: $M=1$~M$_\odot$, $m_1=m_2=1$~M$_\oplus$, $m_3=1$~M$_J$, $a_1=1$~au, $a_3=100$~au, and we set initially $\omega_1=\omega_2=\omega_3=\Omega_1=\Omega_2=\Omega_3=0$,  $e_1=e_2=0.001$, $i_1=i_2=0.001$ and $i_3=45^\circ$. The only difference between each integrated system is the second planet's separation and the third planet's eccentricity. For the {\bf stable} system we chose $a_2=3$~au and $e_3=0.8$. For the {\bf intermediate} system we had $a_2=3$~au and $e_3=0.9$ (which placed it on the stability curve), and finally for the {\bf unstable} system we had $a_2=5.9$~au and $e_3=0.551$. 
}\label{AAA}
\end{figure*}

The proxy also reveals a transitional zone of the parameter space. This zone agrees with our analytically determined zone (i.e., between $e_{3,c}(f_{LL,{\rm min}})$ and $e_{3,c}(f_{LL,{\rm max}})$), where, given the initial conditions in the numerical runs, we set $e_{1,ic}/e_{2,ic}\sim 1$ and $i_{1,ic}/i_{2,ic}\sim 1$. Closer to the stability curve, the orbits might get their eccentricities excited and periodically move closer to one another, but their orbits may never cross. Moreover, far into the top right of Figure 2, the instability will take place sooner in the evolution. In this region, the two inner orbits will undergo EKL evolution independently of each other.

The third planet can excite the inner planets' eccentricities on the EKL timescale. However, eccentricity excitations of each planet will not necessarily cause orbit crossing, as depicted in Figure \ref{fig:sailing}. In the parameter space for our ensemble, we have identified a region containing stable systems that seems to smoothly transition into the instability region. Hence, we identify three regimes: The stable regime, the intermediate regime, and the unstable regime. In the stable regime, the Laplace-Lagrange rapid angular momentum exchange between the two inner planets dominates over the gravitational perturbations of the outer orbit. The left column of Figure \ref{AAA}  depicts this evolution; the system in this column resides in the bottom left of the parameter space in Figure \ref{fig:sailing}. The system remains stable for 10 Gyrs of evolution and the orbits never come close to one another. The two inner planets' inclinations oscillate together around the initial z-axis, due to the precession of the nodes (see below). In the intermediate region, systems might appear to be stable for a long time but the outer orbit perturbations become too dominant, causing instability of the inner system. We show the dynamics of this type of evolution in the middle panel of Figure \ref{AAA}. This system resides very close to the stability criterion plotted in Figure \ref{fig:sailing}, inside the transition zone between $e_{3,c}(f_{LL,{\rm min}})$ and $e_{3,c}(f_{LL,{\rm max}})$. In the unstable regime, gravitational perturbations from the third planet cause high amplitude eccentricity excitation in the inner planets, causing orbits to cross. We show this behavior in the right column \ref{AAA}. This system lies in the top right of the parameter space depicted in Figure \ref{fig:sailing}.

In the system's reference frame (see Figure \ref{fig:Catroon}), the inner planets' inclinations are the angles between their respective angular momenta and the normal to the {\it initial} orbits. Thus, the inclination modulation shown in the stable system (the left column in Figure \ref{AAA}, is due to precession of the nodes, which results in maximum inclination, which is $\sim 2\times i_3$, since we have started with $i_{3,0}=45^\circ$ \citep[e.g.,][]{Inn+97}. 
In contrast, the unstable regime results in misalignment between the inner two planets.

\subsection{Applications: predictions for observed systems }\label{sec:Kepler}

\begin{table}
\begin{center}
\begin{tabular}{|l| c | c | c |}
  \hline
  
 name &  SMA [au]   & mass [M$_J$]   & eccentricity \\
\hline 
\hline
Kepler-36b & 0.12 & 0.015 & $<0.04$ \\
Kepler-36c & 0.13 & 0.027 &$<0.04$ \\
\hline\hline
Kepler-56b & 0.103 & 0.07 & - \\
Kepler-56c & 0.17  & 0.57 & -\\
\hline
\hline
Kepler-88b & 0.097 & 0.027 & 0.06 \\
Kepler-88c & 0.15 & 0.62 & 0.056 \\
\hline\hline
Kepler-109b & 0.07 & 0.023 & 0.21 \\
Kepler-109c & 0.15 & 0.07&0.03\\
 \hline  \hline
Kepler 419b &   0.37     &   2.5           &     0.83         \\
Kepler 419c &   1.68     &     7.3         &     0.18 \\
\hline\hline
Kepler-448b & 	0.15 & 10 & 0.34 \\
Kepler-448c & 4.2 &  22     & 0.65\\
\hline
\end{tabular}
\end{center}
\caption{Observable parameters of the example systems. The observations are adopted from: {\it Kepler-36}: \citet{Carter+12}.  {\it Kepler-56}: \citet{Huber+13} and \citet{Otor+16}.  {\it Kepler-88} \citet{Nesvorny+13}, \citet{Barros+14} and \citet{Delisle+17}. {\it Kepler-109:} \citet{Marcy+14}, \citet{Van+15} and \citet{Silva+15}. {\it Kepler-419}:\citet{Dawson+12}, \citet{Dawson+14} and \citet{Huang+16}. {\it Kepler-448}: \citet{Bourrier+15}, \citet{Johnson+17} and \citet{Masuda17}. 
}\label{table:obs} \vspace{-0.3cm}
\end{table}

\begin{figure*}
\centering
\includegraphics[width=8cm]{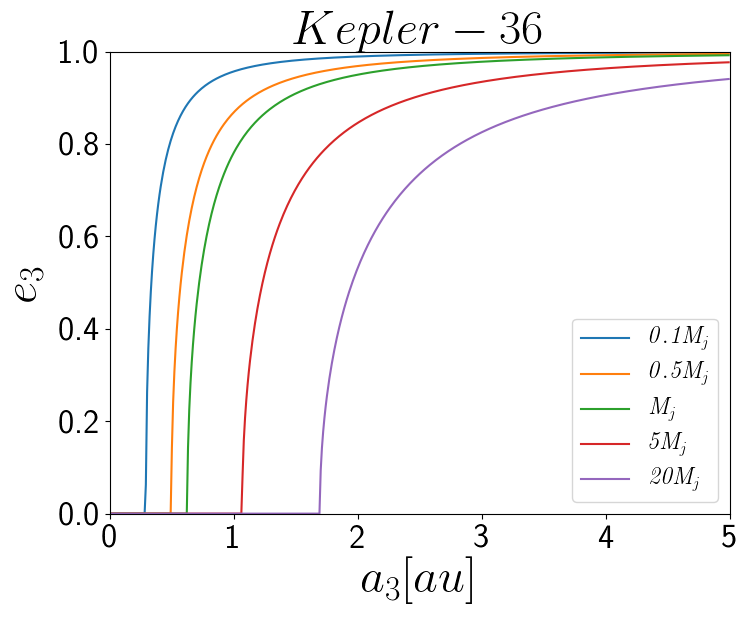}
\includegraphics[width=8cm]{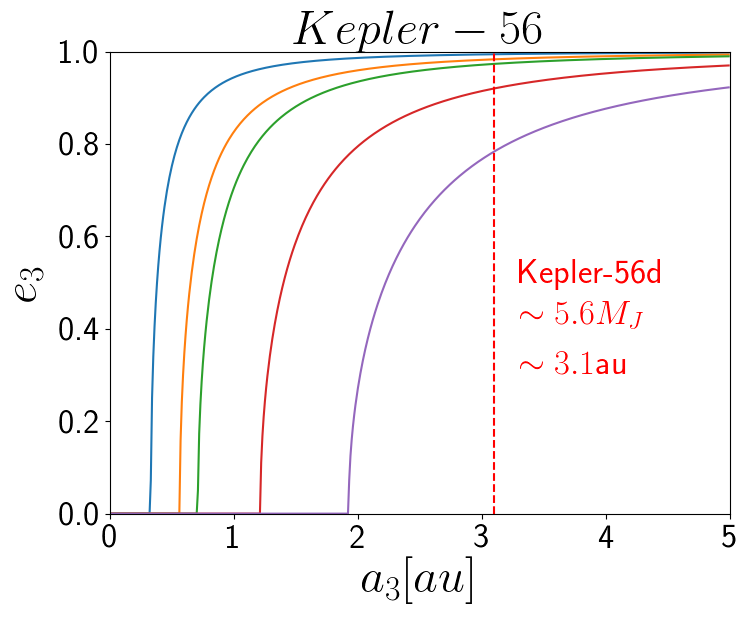}
\includegraphics[width=8cm]{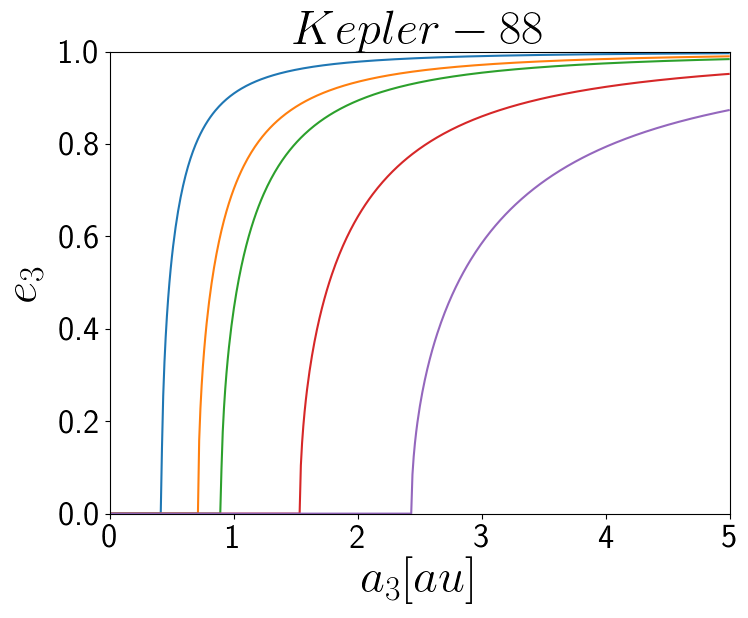}
\includegraphics[width=8cm]{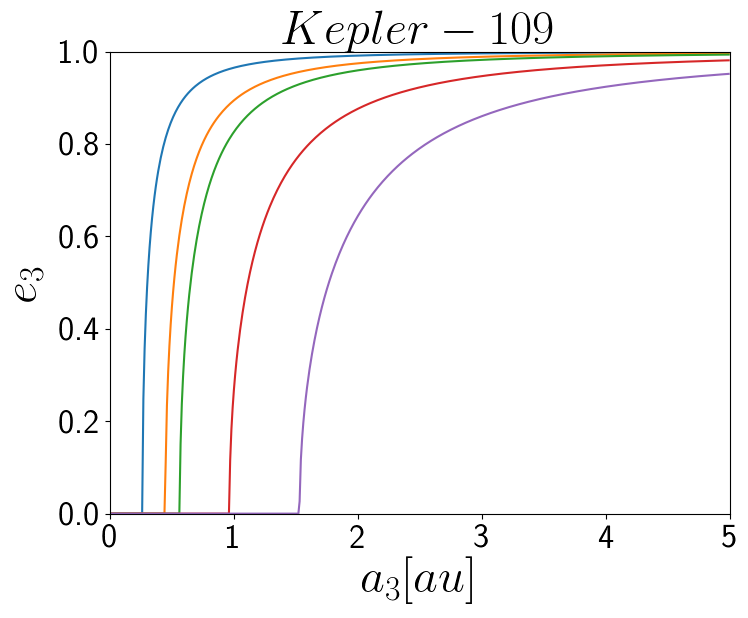}
\includegraphics[width=8cm]{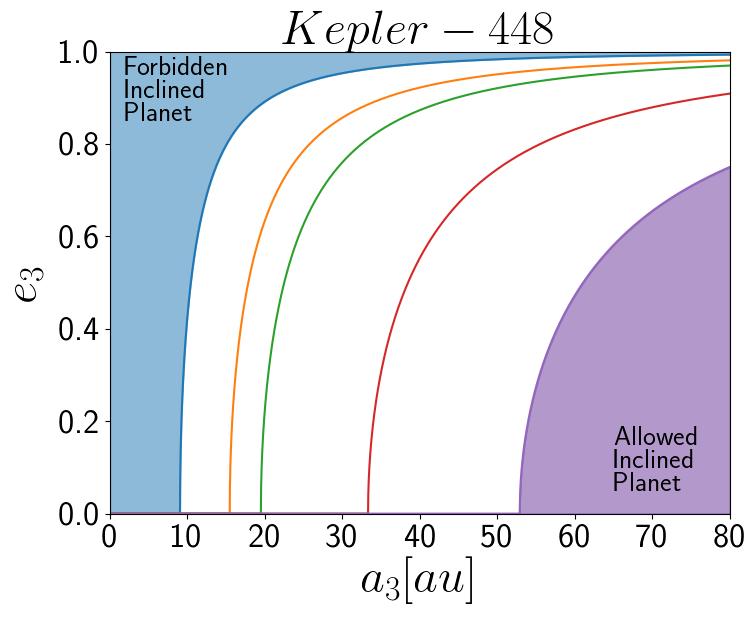}
\caption{{\bf The parameter space of hidden friends for a few observed systems.} We consider the companion's eccentricity $e_3$ and separation $a_3$ for five Kepler systems. For each of the systems we plot the stability criterion $e_{3,c}(f_{LL,{\rm min}})$, for the different companion masses. In particular, we consider companion mass of (from top to bottom), $0.1,0.5,1,5$ and $20$~M$_J$. The stable region exists below {\it each} curve and the instability region resides above each curve (For Kepler-448 we shaded both the stable and unstable regions below the curve corresponding to a $20$~M$_J$ companion and above the curve corresponding to a $0.1$~M$_J$ companion respectively). For each system, we used the observed parameters in Equation \ref{eq:fLLmin} to generate the contours. The observed parameters we used for the inner planets are specified in Table \ref{table:obs}. We note that a third companion was reported to Kepler-56, with a minimum mass of $5.6$~M$_J$, which yields a $3.1$~au separation \citep{Otor+16}. This constrains the eccentricity of the companion to lie on a vertical line of a constant semi-major axis in the parameter space. As such, we over-plotted Kepler-56d on the top right panel (dashed line).   }
	\label{fig:Kepler}\vspace{0.6cm}

\end{figure*}

The stability criterion derived here can be used to predict the parameter space in which a hidden companion can exist within a system without destabilizing it. As a proof-of-concept, we discuss the stability of a few non-resonant observed exoplanetary systems in the presence of inclined planets. Specifically, for some observed systems, we provide a set of predictions that characterize the ranges of parameter space available for hidden companions to exist in without disrupting inner orbits. We focus on the following systems: Kepler-419, Kepler-56, Kepler-448, and Kepler-36. These represent the few systems that characterize the extreme limits of two planet systems, from tightly packed super-earths such as Kepler-36 to hierarchical eccentric warm Jupiter systems such as Kepler-448. In Figure \ref{fig:Kepler} we show the relevant parameter space in which the systems can have a hidden inclined companion and remain stable. We chose the more conservative stability criterion for this exercise, i.e.,  $e_{3,c}(f_{LL,{\rm min}})$. Each line in the four panels of Figure \ref{fig:Kepler} shows the stability criterion for different companion masses. In particular,  we consider companion masses of (from top to bottom), $0.1,0.5,1,5$ and $20$~M$_\odot$. For each companion mass, allowed system configurations lie below the curve, and unstable configurations are above the curve.  We caution that very close to the stability curve (even below the curve) systems may still undergo eccentricity excitations that may affect the dynamics. The parameters of the inner, observed planets were taken from observations, see Table \ref{table:obs}.  Below we discuss the specifics of the four example systems. 

\begin{itemize}
\item {\it Kepler-36} is an, approximately, one solar mass star orbited by two few-Earth mass planets on $13.8$ and $16.2$~days orbits \citep{Carter+12}. This compact configuration is expected to be stable in the presence of perturbations as shown in Figure \ref{fig:Kepler}, in the top left panel. For example, as can be seen in the Figure, an eccentric inclined $20$~M$_J$ brown dwarf can reside slightly beyond $2$~au.
\item {\it Kepler-56} is an evolved star ($M=1.37$~M$_\odot$) at the base of the red giant branch and is orbited by two sub-Jovian planets with low mutual inclinations  \citep[e.g.,][]{Huber+13}. Most notably, asteroseismology analysis placed a lower limit on the two planets' obliquities of $37^\circ$. \citet{Li+14Kepler} showed that a large obliquity for this system is consistent with a dynamical origin. They suggested that a third outer companion is expected to reside far away and used the inner two planets' obliquities to constrain their inclinations. Follow-up radial velocity measurements estimated that a third companion indeed exists with a period of $1000$~days and a minimum mass of $5.6$~M$_J$ \citep[e.g.,][]{Otor+16}. Here we show that indeed a $\sim 5$~M$_J$ planet can exist at $\sim 3$~au, with a range of possible eccentricities up to $0.9$. A more massive planet can still exist at $\sim 3$~au with a possible range of eccentricities up to slightly below $e_3\sim 0.8$. This example of a tightly packed system yields the expected result, i.e, a large part of the parameter space is allowed to have an inclined companion, as depicted in Figure \ref{fig:Kepler} in the top right panel. In the presence of an inclined companion, the inner two planets will most likely have non-negligible obliquities, as was postulated by \citet{Li+14Kepler}. 
\item {\it Kepler-88} is a star of, approximately, one solar mass. It has been observed to be orbited by two planets. The first planet is Earth-like ($m_1\sim 0.85M_{\oplus}$) and the second planet is comparable to Jupiter ($m_2\sim0.67$~M$_{J}$). Together, the two planets form a compact inner system with negligible eccentricities \citep[e.g.,][]{Nesvorny+13,Barros+14,Delisle+17}. 
In Figure \ref{fig:Kepler}, left middle panel, we show that an inclined planet can exist in large parts of the parameter space. For example, a massive companion ($\sim 20$~M$_J$) can exist beyond $2$au with an eccentric orbit ($\sim 0.7$). 
\item {\it Kepler-109} is a star that also has, approximately, one solar mass ($1.07$~M$_\odot$) and two planets orbiting it in a compact configuration, having a small mutual inclination \citep{Marcy+14,Van+15,Silva+15}. However, the eccentricity of the innermost planet is not as negligible as the secondary's ($e_1 \sim 0.21 > e_2 \sim 0.03$), but does not yield a violation of the approximation. We show the stability bounds within the $(e_3,a_3)$ parameter space for this system in the middle right panel of Figure \ref{fig:Kepler}. There it can be seen that this system can exist in the presence of an eccentric companion with $20$~M$_J$, so long as the companion is beyond $3$au.
\item {\it Kepler-419}  is a $1.39$~M$_\odot$  mass star which is orbited by two Jupiter sized planets with non-negligible eccentricities and low mutual inclinations \citep[e.g.,][]{Dawson+12,Dawson+14,Huang+16}. The large, observed eccentricities of the planets violate the assumption that the eccentricities are sufficiently small. Recall that this assumption was used to derive the stability criterion in Eq.~(\ref{eq:e3c}). Furthermore, the quadrupole timescale for precession induced by Kepler-419c on Kepler-419b is comparable to that of Laplace-Lagrange's. Thus, a distant massive companion is expected to excite the two planets' eccentricities and inclinations. When the mutual inclination between the two inner planets is large, the second planet can further excite the innermost planets' eccentricity, thus rendering the system unstable. We have verified, numerically, the instability of the system is consistent with the breakdown of our criterion. We did this for several systems with far away companions with masses varying from $1,5,$ and $20$~M$_J$. Thus, we suggest that a massive inclined companion can probably be ruled out for this system. Since our stability criterion is violated here, it is not necessary to show it's parameter space in a plot. On the other hand, a smaller companion is not expected to cause secular excitations in the eccentricities of more massive planets, since the inner system will excite its eccentricity and inclination \citep[e.g., inverse EKL][]{Naoz+17,Zanardi+17,Vinson+18}. Therefore, the existence of a smaller inclined companion cannot be ruled out.  
\item Kepler-448  is a $1.45$~M$_\odot$ star orbited by a $10$~M$_J$ warm Jupiter  \citep{Bourrier+15,Johnson+17}. Recently, there was a reported discovery of a massive companion ($\sim 22$~M$_J$) with a rather hierarchical configuration  \citep{Masuda17}. The hierarchical nature of the inner system yields a more limited range of separations to hide a companion, see Figure \ref{fig:Kepler}, in the bottom panel. On the other hand, the large masses of Kepler-448b and Kepler-448c imply that a small planetary inclined companion can still exist with negligible implications on the inner system. In fact, one would expect that the inner system will largely affect a less massive companion \citep[e.g.,][]{Naoz+17,Zanardi+17,Vinson+18}.
\end{itemize}

\section{discussion}\label{sec:dis} 

We have analyzed the stability of four-body hierarchical systems, where the forth companion is set far away ($a_3\gg a_2,a_1$). Specifically, we focus on three planet systems, for which the two inner planets reside in a relatively close configuration and have an inclined, far away, companion. Observations have shown that multiple tightly packed planetary configurations are abundant in our Galaxy. These systems may host a far away companion, which might be inclined or even eccentric. An inclined perturber can excite the eccentricity of planets in the inner system via the EKL mechanism, which can ultimately destabilize the system.

We have analytically determined a stability criterion for two-planet systems in the presence of an inclined companion (Equation \ref{eq:e3c}).  This criterion depends on the initial conditions of the inner planetary system and on the outer orbit's mass, eccentricity, and separation. 
It is thus straightforward to generalize it to $n>2$ inner planetary systems. 
We then numerically integrated a set of similar systems using the Gauss's averaging method, varying only the outer companion's eccentricity $e_3$ and the second planet's separation $a_2$. We have 
characterized each numerical integrated systems' stability and showed that stable systems are consistent with our analytical criterion.

A system will remain stable if the timescale for angular momentum exchange between the two inner orbits takes place faster than eccentricity excitations that might be induced by a far away, inclined, companion. When the system is stable, the two inner orbits have minimal eccentricity modulations and their inclinations, with respect to their initial normal orbit, remained aligned to one another. An example of such as system is depicted in the left panels of Figure \ref{AAA}.  

Assuming the normals of the inner orbits are initially parallel to the stellar axis allows precession of the nodes to be interpreted as obliquity variations. Thus, a non-negligible obliquity for two (or more) tightly packed inner orbits may be a signature of an inclined, distant, companion \citep[e.g.,][]{Li+14Kepler}. The obliquity, in this case, oscillates during the system's dynamical evolution, which can have large implications on the habitability of the planets  
\citep[e.g.,][]{Shan+17}.

On the other hand, a system will destabilize if the precession induced by the outer companion is faster than the precessions caused by interactions between the inner bodies. In this case, the two inner planets will exhibit large eccentricity excitations accompanied with large inclination oscillations (see, for example, right panels in Figure \ref{AAA}). In this type of system, each planet undergoes nearly independent EKL oscillations, and thus extremely large eccentricity values can be expected, as well as chaos \citep[e.g.,][]{Naoz+13,Teyssandier+13,Li+14}.

We also showed (e.g., Figure \ref{fig:sailing}, green band) that the stability criterion includes a transition zone, where systems are likely to develop large eccentricity, leading to close orbits. Systems close to the transition zone, or in the transition zone, can be stable for long period of time, and develop instability very late in the evolution. We show an example for such as system in the middle column of Figure \ref{fig:sailing}. In this example, the system stayed stable for slightly more than a Gyr and developed an instability that leads to orbit crossing after $\sim 1.5$~Gyr. 

 We note that our analysis did not include General Relativistic or tidal effects between the planets and the star because, typically, including them will further stabilize systems. General relativistic precession tends to suppress eccentricity excitations if the precession takes place on a shorter timescale than the induced gravitational perturbations from a companion   \citep[e.g.,][]{Naoz+12GR}. Also, tidal precession tends to suppress eccentricity excitations \citep{Dan,Liu+15}. These suppression effects become more prominent the closer the inner orbits are to the host star. Thus, if eccentricity excitations from the companion take place on a longer timescale than general relativity or tidal precession the system will remain stable.  

Finally, as a proof-of-concept, we used our stability criterion to predict the parameter space in which a hidden inclined companion can reside for four Kepler systems (see Figure \ref{fig:Kepler}). The systems we consider were Kepler-419, Kepler-56, Kepler-448, and Kepler-36. These systems represent a range of configurations, from tightly packed systems with small or super-Earth mass planets to potentially hierarchical systems with Jupiter mass planets. A notable example is Kepler-56, where the recently detected third planet was reported to have a minimum mass of $5.6$~M$_J$, and a $\sim 1000$~days orbit. Such a system indeed resides in the predicted stable regime. Furthermore, given a mass for the Kepler-56d, we can limit its possible eccentricity.  

\acknowledgements
PD acknowledges the partial support of the Weyl Undergraduate Scholarship. SN acknowledges the partial support of the Sloan fellowship. 

 \bibliographystyle{apj}
 \bibliography{Kozai}

\end{document}